\documentclass[twocolumn]{aa}
\usepackage{graphicx}
\usepackage{txfonts}
\begin{document}
   \title{Interstellar extinction in the direction
   of the Aquila Rift}

   \author{V. Strai\v zys\inst{1}
          \and
          K. \v Cernis\inst{1}
          \and
          S. Barta\v si\=ut\.e\inst{1,2}
          }

   \offprints{V. Strai\v zys}

   \institute{Institute of Theoretical Physics and Astronomy,
             Vilnius University, Go\v stauto 12, Vilnius 2600,
Lithuania\\
             \email{straizys@itpa.lt; cernis@itpa.lt}
         \and
             Astronomical Observatory of Vilnius University,
             \v Ciurlionio 29, Vilnius 2009, Lithuania\\
             \email{stanislava.bartasiute@ff.vu.lt}
             }

    \date{Received February 15, 2003; accepted March 15, 2003}

   \abstract{
   The distance dependence of interstellar extinction in the
   direction of the \object{Aquila Rift} is investigated using 473 stars
   observed in the Vilnius photometric system.  Front edge of the
   dark clouds in the area is found to be at $225\pm55$ pc and the
   thickness of the cloud system can be about 80 pc.  The maximum
   extinction $A_V$ in the clouds is close to 3.0 mag.  Two stars
   with larger extinction are found and discussed.  Since the new
   distance of the clouds is larger than the previously accepted distance,
   the cloud system mass should be increased up to $2.7\times10^5
   M_{\odot}$ which is close to the virial mass estimated from the CO
   velocity dispersion.  Additional arguments are given in favor of
   the genetic relation between the Serpens and the Scorpio-Ophiuchus
   dark clouds.
    }

  \titlerunning{Interstellar extinction in the
   Aquila Rift}
  \authorrunning{V. Strai\v zys et al.}

   \maketitle

\section{Introduction}

Starting from Cygnus, in the direction of the Galactic center the
Milky Way appears to split into two branches.  The southern branch
runs through Cygnus, Vulpecula, Sagitta, Aquila and Scutum,
entering the Galactic central bulge in Sagittarius.  The northern
branch crosses Vulpecula and Aquila and disappears in the northern
part of the Serpens Cauda and Ophiuchus constellations, being
covered by numerous dust clouds.  This complex of dark clouds
usually is called the \object{Aquila Rift}.

The distances and extinction properties of these clouds are known
only with low accuracy. So far the area is poorly investigated by
modern photometric methods in the optical range.  The collected
{\it UBV} and MK data have been used for a crude estimate of the
dependence of interstellar extinction on distance in the Rift
direction (FitzGerald \cite{fitzg}, Neckel \& Klare \cite{neckel},
Forbes \cite{forbes} and others). A sudden appearance of reddened
stars at 200--250 pc is observed.  According to the summary of
Dame \& Thaddeus (\cite{dame85}), the estimated distance of the
\object{Aquila Rift} is 200$\pm$100 pc.

A field around the core of one of the densest Serpens molecular
clouds (at 18$^{\rm h}$30$^{\rm m}$, +1$\degr$14.5$\arcmin$,
2000.0) with active star formation has attracted more attention
(see the review article by Eiroa \cite{eiroa}).  The distance of
the cloud has been discussed by Strom et al. (\cite{strom}),
Chavarria et al. (\cite{chavar87}, \cite{chavar88}), Zhang et al.
(\cite{zhang}), de Lara \& Chavarria (\cite{lara89}), de Lara et
al.  (\cite{lara91}) and Strai\v zys et al.  (\cite{strai96}).

Stars and other sources in the area have been well covered by the
infrared surveys:  by IRAS in the far infrared and by 2MASS in the
$JHK$ range.  Also, numerous radioastronomical studies in the
lines of H\ts I, CO, H$_2$CO, NH$_3$ and H$_2$O are available in
the area.  The distribution of molecules, especially of CO, shows
a very close resemblance to the dust distribution (see Dame \&
Thaddeus \cite{dame85}; Dame et al.  \cite{dame87},
\cite{dame01}).  According to CO radio observations, the
\object{Aquila Rift} occupies a region of irregular form between
20$\degr$ and 40$\degr$ in Galactic longitude and between
--6$\degr$ and +14$\degr$ in Galactic latitude.  However, some
protrusions and blobs of gas and dust extend up to $\ell$ =
15$\degr$ and $b$ = +20$\degr$.

About a decade ago we started a program of investigation of the Serpens
Cauda clouds belonging to the Aquila Rift by photoelectric photometry of
stars in the {\it Vilnius} seven-color photometric system
(Table~\ref{VilnBands}).  Our first results were published by Strai\v
zys, \v Cernis \& Barta\v si\= ut\.e (\cite{strai96}), hereafter Paper
I. The investigation was based on photometry and photometric
classification of 105 stars down to magnitude 13, located around the
core of the Serpens molecular cloud mentioned above.  The size of the
area investigated was about 6.5 square degrees.  It was found that the
dust cloud in the area appears at a distance of about 260 pc.

\begin{table}
      \caption[]{ Mean wavelengths and half-widths of passbands
                 of the Vilnius photometric system.}
          \label{VilnBands}
       $$
          \begin{array}{p{0.27\linewidth}rrrrrrr}
            \hline
            \noalign{\smallskip}
            Passband & U~ & P~ & X~ & Y~ & Z~ & V~ & S~ \\
            \noalign{\smallskip}
            \hline
            \noalign{\smallskip}
            ~~$\lambda$ (nm) & ~~345 & ~~374 & ~~405 & ~~466 &
                             ~~516 & ~~544 & ~~656 \\
            $\Delta\lambda$ (nm) & 40 & 26 & 22 & 26 & 21 & 26 & 20 \\
            \noalign{\smallskip}
            \hline
          \end{array}
       $$
   \end{table}

      \begin{figure}
   \resizebox{\hsize}{!}{\includegraphics{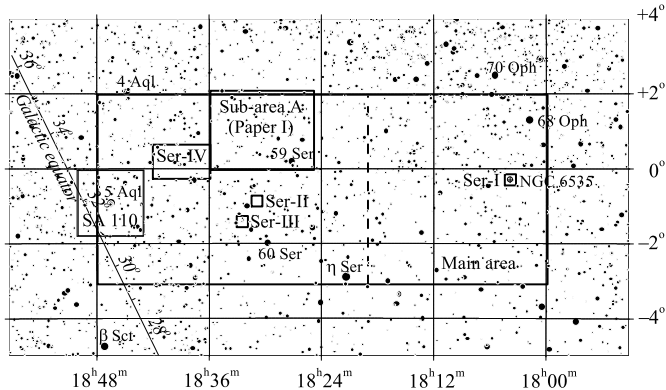}}
      \caption{
      ~The chart for the investigated Serpens Cauda area with the
      following sub-areas: A is the sub-area investigated in Paper I;
      Ser I, Ser II, Ser III and Ser IV are the sub-areas in which
      photoelectric standards have been measured for future CCD
      photometry (\v Cernis et al. \cite{cernis} and Paper III);
      \object{SA 110} is the area in which HD stars have been measured by
      Zdanavi\v cius et al. (\cite{zdan}). The broken line at
      RA=18$^{\rm h}$49$^{\rm m}$ divides the area into two parts with
      different dependence of $A_V$ on distance.
                    }
         \label{Map}
   \end{figure}

      \begin{table}
      \caption[]{ Four sub-areas in which fainter stars have
                  been measured in the Vilnius photometric system
                  and SA 110.}
          \label{VilnObs}
       $$
          \begin{array}{p{0.2\linewidth}ccrcc}
            \hline
            \noalign{\smallskip}
            Sub- & {\mathrm{RA}} & ~{\mathrm{DEC}} & {\rm
            Magnitude}
            & {\rm Number} & {\rm Publ.} \\
            {\rm area} & {\mathrm{h~~~~m}} & ~{\scriptstyle{\circ~~~~\prime}}  &
            {\rm limits~~~} & {\rm of~stars} & \\
            \noalign{\smallskip}
            \hline
            \noalign{\smallskip}
            Ser~I & 18~~01.3 & -00~~18~~ & 11.3-14.4 & 11 & 1 \\
           Ser~II & 18~~31.5 & -00~~50~~ & 11.0-13.4 & ~4 & 2 \\
          Ser~III & 18~~32.5 & -01~~23~~ & 10.2-12.5 & ~7 & 2 \\
           Ser~IV & 18~~39.0 & +00~~10~~ &  8.0-13.3 & 45 & 2 \\
           SA~110 & 18~~46.5 & -01~~00~~ &  6.9-10.7 & 30 & 3 \\
            \noalign{\smallskip}
            \hline
          \end{array}
       $$
       \begin{list}{}{}
       \item Publications: (1) \v Cernis et al. (\cite{cernis}), (2)
        Strai\v zys et al. (\cite{strai02c}, Paper III),
        (3) Zdanavi\v cius et al. (\cite{zdan}).
       \end{list}
   \end{table}

However, the area investigated in Paper I is only a small part of the
whole complex of the \object{Aquila Rift}.  The newest catalog of dark
clouds of Dutra \& Bica (\cite{dutra}) enumerates in the Rift more than
50 clouds of different sizes.  It is important to know whether all these
clouds are at the same distance or they form a system with a significant
depth.

In the present paper we investigate interstellar extinction and
cloud distances in a much larger area, covering 5$\times$10 sq.
degrees.  The area is limited by the following 2000.0 coordinates:
RA from 18$^{\rm h}$00$^{\rm m}$ to 18$^{\rm h}$48$^{\rm m}$ and
DEC from --3.0$\degr$ to +2.0$\degr$. Photometry in the {\it
Vilnius} system was obtained in 1994, 1997 and 2001 with the 1
meter telescope at the Maidanak Observatory in Uzbekistan.  The
results of photometry of 419 stars down to 11 mag and their
photometric classification were published by Strai\v zys, \v
Cernis \& Barta\v si\= ut\.e (\cite{strai02b}), hereafter Paper
II.

Additionally, 67 fainter stars with $V$ between 10th and 14th mag
were observed in four smaller sub-areas situated within the large
area indicated above.  These sub-areas are listed in
Table~\ref{VilnObs}. The measured stars in the sub-areas I--IV
will be used as standards for future CCD photometry of fainter
stars. Kapteyn Selected Area 110 is located at the edge of the
same area. In it {\it Vilnius} photometry of 30 HD stars has been
published by Zdanavi\v cius et al. (\cite{zdan}).  These stars
were also included in the present study of interstellar extinction
(see Table~\ref{VilnObs}). Magnitudes and color indices of stars
in the sub-areas are published in the papers listed in
Table~\ref{VilnObs}.

In the present study the extinction data of 80 stars from Paper I were
also used.  Their distances are transformed to the new distance scale
corresponding to the distance modulus of \object{Hyades} $V-M_V=3.3$.
Other stars of Paper I were rejected:  some of them were found to be
visual binaries and for some the classification accuracy from the
photometric data was too low.  They are either unresolved binaries or
peculiar objects.

Consequently, we had at our disposal photometry of about 600 stars
in total (14 stars are common to the catalogs of Papers I and II
and 8 stars are common to the catalogs of Paper II and SA 110).
However, for the investigation of extinction in the area we used
473 stars only, after the exclusion of binary, multiple and
peculiar stars.  The map of the investigated area with the
sub-areas is shown in Fig.~\ref{Map}.

\section{Interstellar extinction law}

We have identified 43 stars from our list with the 2MASS survey
photometry available through the Internet (Skrutskie et al.
\cite{skrutsk}).  For 19 stars, covering the range of extinctions
$A_V$ from 0.4 to 2.5 mag, we have calculated color indices $V-K$
taking $V$ from Table~1 of Paper II, and color excesses $E_{V-K}$
taking the intrinsic $(V-K)_0$ from Strai\v zys (\cite{strai92},
Tables 22--24). Spectral types of these stars were taken from
spectroscopic classification.  The least squares solution gives
the equation:
   \begin{equation}
    E_{V-K}/E_{Y-V} = 3.295 + 0.347\,(Y-V)_0 \pm 0.464\,.
    \label{Eratio}
   \end{equation}

This equation shows that the ratio $E_{V-K}/E_{Y-V}$ varies from
$\sim\,$3.4 for A-type stars to $\sim\,$3.6 for K giants.  Since
the ratio $E_{Y-V}/E_{B-V} \approx 0.8$ for A stars and
$\approx\,$0.85 for K giants, Eqn.~(\ref{Eratio}) leads to
$E_{V-K}/E_{B-V}$ between 2.7 and 3.1 and $R_{BV} =
1.1\,E_{V-K}/E_{B-V}$ between 3.0 and 3.4.  These values of
$R_{BV}$ are very close to the ratio given by the normal
interstellar extinction law (Strai\v zys \cite{strai92}).

According to Cardelli et al.  (\cite{cardel88}, \cite{cardel89}),
the form of the extinction law in the visible and the infrared
ranges, defining the ratio $R_{BV}$, is well correlated with its
form in the ultraviolet.  Thus, we may accept that in the Serpens
dark clouds of medium density the extinction law is normal, i.e.,
typical for the diffuse dust.  This is expected, since the area
does not contain young hot and luminous stars which may modify
grain sizes in the interstellar medium.  Only in the dense core of
the Serpens molecular cloud, in the vicinity of some B-type stars
embedded in the cloud, Chavarria et al. (\cite{chavar88}) and de
Lara et al.  (\cite{lara91}) have found a larger than normal
$R_{BV}$ ratio.  However, in other parts such interaction between
the hot stars and interstellar dust is not observed. Thus, in the
calculation of the reddening-free $Q$-parameters we used the
ratios of color excesses corresponding to the normal extinction
law.

   \section{Quantification of stars and their interstellar reddening}

At the beginning all known and suspected binary stars among the observed
stars were identified and excluded from further investigation of
interstellar extinction.  For this we have identified visual binaries in
the Washington Double Star Catalog (Mason et al.  \cite{mason}).  A
number of stars were suspected as binaries by inspecting their images in
Internet's virtual telescope SkyView of NASA based on the Digital Sky
Survey scans of the Palomar atlas plates (http://skyview.gsfc.nasa.gov).
About 8\% of stars were suspected as binary or multiple stars since
their images were found to be non-symmetrical.  In some cases close
satellite stars were detected:  they also might affect photometry of the
main component.  The real and suspected binaries are identified in the
catalogs of Papers II and III.  The suspected binaries in the area of
Paper I will be listed below.

For the determination of spectral classes and absolute magnitudes of
stars from color indices we used the interstellar reddening-free
$Q$-parameters defined by the equation:
   \begin{equation}
           Q_{1234} = (m_1-m_2) - (E_{12}/E_{34})(m_3-m_4)\,,
   \end{equation}
where
   \begin{equation}
           E_{k,\ell} = (m_k-m_{\ell})_{\rm {reddened}} -
                        (m_k-m_{\ell})_{\rm {intrinsic}}\,.
   \end{equation}
Two independent methods described in Strai\v zys et al.
(\cite{strai01b}) were applied:  (1) the $\sigma Q$ method which
uses matching of 14 different reddening-free $Q$ parameters of a
program star to those of about 8400 stars with known spectral and
luminosity classes in the MK system, metallicities and
peculiarities and (2) the $Q,Q$ method which uses interstellar
reddening-free $Q,Q$ diagrams calibrated in spectral classes and
absolute magnitudes.  The last method was used only for G5--K--M
stars (the $Q_{UPY}$,$Q_{XZS}$ and $Q_{XZS}$ $Q_{XYZ}$ diagrams.
For spectral classes earlier than G5 only the $\sigma Q$ method
was used.  For each program star three closest MK stars were
selected.  Absolute magnitudes were taken from the MK type
tabulation of Strai\v zys (\cite{strai92}).

The results of photometric quantification of stars and determination of
their interstellar extinctions and distances are given in Paper III
(Strai\v zys, Barta\v si\= ut\.e \& \v Cernis \cite{strai02c}).  For the
transformation of color excesses to interstellar extinctions the normal
value of the ratio $R_{YV} = A_V/E_{Y-V} = 4.16$ was used (Strai\v zys
et al.  \cite{strai96}).  For the stars in the densest parts of the
Serpens molecular cloud the larger values of $R_{YV}$ were used (see
Section 2).  The distances $r$ of the stars were calculated by the
equation
   \begin{equation}
          \log r = {{V-A_V+5-A_V}\over 5}\,.
   \end{equation}
The expected errors are:  $\pm$0.03 mag for $E_{Y-V}$, $\pm$0.1
mag for $A_V$ and $\pm$25\% for distance.

Papers I and III contain 38 stars closer than 250 pc with
photometric distances and Hipparcos parallaxes available.
Differences of the photometric and trigonometric distances of 32
stars fall within the $\pm$25\% limits, as expected from the
accuracy of the photometrically determined absolute magnitudes and
the parallax errors.  Larger differences for the remaining six
stars may be explained by the photometric luminosity errors caused
by undetected duplicity or peculiarity.  These stars are not of
decisive importance in the future study of interstellar extinction
in the area.  A similar comparison of photometric and
trigonometric distances was done earlier in the \object{California
nebula} region (Strai\v zys et al.  \cite{strai01a}) and in the
Aries molecular cloud region (Strai\v zys et al. \cite{strai02a}).
In all cases the agreement of distance scales was found to be
sufficiently good, with no systematic differences.

\section{Interstellar extinction versus distance}

\begin{figure}
    \resizebox{\hsize}{!}{\includegraphics{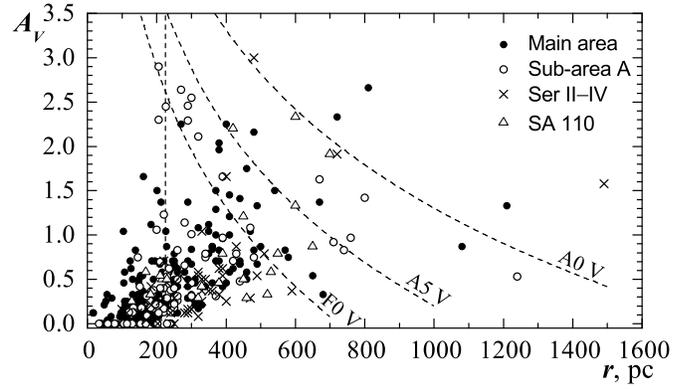}}
       \caption{Interstellar extinction $A_V$ plotted against distance
                $r$ in parsecs for the east part of the Serpens Cauda
                area with RA between 18$^{\rm h}$19$^{\rm m}$ and
                18$^{\rm h}$48$^{\rm m}$. The vertical broken line
                shows the accepted distance of the front dust
                layer at 225 pc.  The broken curves show the limiting
                magnitude effect for A0\ts V, A5\ts V and F0\ts V stars of
                $V$ = 12 mag.
              }
          \label{Aeast}
   \end{figure}

\begin{figure}
     \resizebox{\hsize}{!}{\includegraphics{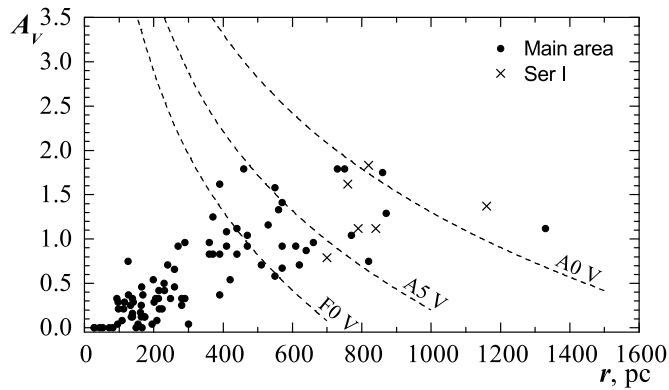}}
       \caption{Interstellar extinction $A_V$ plotted against
       distance $r$ in parsecs for the west part of the Serpens
       Cauda area with RA between 18$^{\rm h}$00$^{\rm m}$ and
       18$^{\rm h}$19$^{\rm m}$. The limiting magnitude curves are
       described in Fig.~\ref{Aeast}.
              }
          \label{Awest}
   \end{figure}

The extinction $A_V$ for stars in the investigated area are
plotted vs. distance in Figs.  \ref{Aeast} and \ref{Awest} for two
parts of the area. The division line is at RA=18$^{\rm h}$49$^{\rm
m}$, which separates the western part (from the Galactic equator
to $b$ $\sim$ 7$\degr$) with heavier extinction and the eastern
part ($b$ $\sim$ 7--11$\degr$) with smaller extinction. On both
Figures the stars from the main area (Paper II) are plotted as
dots, the stars from the molecular cloud area (Paper I) -- as
circles, the stars from Areas I to IV -- as crosses, the stars
from \object{SA 110} -- as triangles.

Fig.~\ref{Aeast} shows that the majority of stars in the graph
form a wedge-shaped belt, which starts at the origin of the
coordinates. The stars with nearly zero-reddening extend from the
Solar vicinity up to $\sim$300 pc.  The upper limit of the wedge
grows up with the distance and at 160 pc $A_V$ reaches $\sim$0.8
mag. Only one star, BD--2 4634 (or No. 31 in Paper II) shows an
anomalous value of extinction:  at 104 pc $A_V$ is 1.04.  It is
not excluded that the star may be an unresolved binary.  At 160 pc
a sudden increase of $A_V$ takes place.  The majority of stars
continue to follow the wedge-shaped pattern, however, a smaller
number of stars exhibit much larger values of $A_V$:  up to 1--3
mag and even more.  This is an indication that somewhere between
200 and 400 pc a network of dense dust clouds starts to appear.
However, these clouds do not cover all the area since many stars
at these distances show only moderate extinction, between 0.4 and
1.0 mag.

Let us estimate the distance of the nearest dust concentrations.
The standard deviation of distances is expected to be about $\pm
0.25r$. This means that the apparently closest heavily reddened
stars may appear at a distance of $r-0.25r=160$ pc, i.e.,
$r=160/0.75=213$ pc.  Another estimation of the front edge
distance of the clouds comes from the unreddened or slightly
reddened stars, exhibiting the largest apparent distances.  For
them $r+0.25r=300$ pc, i.e.,  $r=300/1.25=240$ pc.  Thus, the real
distance of the front dark cloud, responsible for the rise of
extinction at $r>160$ pc, is between 213 and 240 pc, i.e.
approximately at $225\pm 56$ pc.  It is interesting that the
stars, exhibiting heavy extinction within this distance range do
not show any correlation with the apparent richness of the
background stars.

In Fig.~\ref{Awest}, which shows the stars at higher distances
from the Galactic plane, the extinction up to 100 pc is nearly
zero. Beyond this distance, the extinction increases gradually up
to 300--400 pc and the maximum $A_V$ values do not exceed 1.8 mag.
This means that the extinction in the area is much lower and more
uniform in comparison with lower Galactic latitudes.

There is no possibility to estimate the largest extinction value
at different distances, since in Figs.  \ref{Aeast} and
\ref{Awest} a strong selection effect is present.  The dotted
lines show the limiting magnitude effect at $V$ = 12 mag for the
stars of spectral classes A0\ts V, A5\ts V and F0\ts V. The stars
of these spectral types above the corresponding curves are
accessible only if the $V$ magnitude limit is set at $V > 12$. The
limiting magnitude in various directions of the area varies from
10 to 13. Consequently, the distribution of stars in the figures
is strongly affected by selection, and the number of heavily
reddened stars at distances $>400$ pc is considerably reduced.
Therefore, we are not able to give any maximum or medium
extinction values at these distances. Moreover, a number of more
distant dust clouds may be present. However, the lowest extinction
value in the area can be estimated with sufficiently high
accuracy.  At distances $r>400$ pc it is about 0.5 mag.  The stars
with such low extinction are seen through windows between the dust
clouds.

The star \object{BD --01 3542} ($V = 9.24$) deserves a special
discussion.  According to Hiltner \& Iriarte (\cite{hiltner}) its
spectral type is B8\ts Ia:.  Nassau \& Stephenson (\cite{nassau})
in their Catalog of Luminous Stars in the Northern Milky Way list
this star as \object{LS\ts IV -01 4} with spectral type OB+R where
R means ``reddened".  Bidelman (\cite{bidel}) finds in its
spectrum a ``very weak broad H$\alpha$ emission".  Our photometric
classification gives a spectral class B2/3 and a luminosity
II--III (uncertain).  If the star were B3\ts II--III, its $A_V$
should be 5.0, $M_V\approx -4.0$ and the distance 440 pc.  It is
unlikely that so heavily reddened star is at this small distance
and in the area with a relatively rich Milky Way background.  If
we accept the B8~Ia type, its $A_V$ becomes 4.6 mag, $M_V\approx
-7.3$ mag and the distance 2.4 kpc.  This combination of $A_V$ and
$r$ seems to be more acceptable, since the star is only at
+1$\degr$ latitude. The star deserves careful spectral and
photometric investigation.

Another interesting star from our list is \object{BD --02 4676},
with $V = 10.25$.  This is the irregular Lb-type variable
\object{CZ Ser} of spectral class M6\ts III (photometric
classification). The same spectral class is given by Hanson \&
Blanco (\cite{hanson}) from low-dispersion objective-prism
spectra.  The General Catalogue of Variable Stars (Kholopov
\cite{kholopov}) and Sloan \& Price (\cite{sloan}) list a spectral
class of M6.5 for the star. The star is an infrared source
\object{IRAS 18347-0241} and it has a circumstellar silicate and
CO envelope (Sloan \& Price \cite{sloan}; Kerschbaum \& Olofsson
\cite{kersch}).  We obtain its reddening $E_{Y-V}\approx 0.84$. If
all this reddening is of interstellar origin, its $A_V = 3.5$ mag,
and with $M_V = 0.0$ mag a distance of 225 pc follows. However,
since part of the reddening is circumstellar, its $A_V$ and
distance are very uncertain (see also a discussion of Olofsson et
al. (\cite{olof})).

   \section{Discussion and conclusions}

The investigations described in Paper I and the present paper give
evidence that the dust clouds of the \object{Aquila Rift} begin at the
200--250 pc distance.  These clouds cover the whole investigated area
including regions with different richness of the background stars, seen
on the red Palomar Sky Survey plates.  This means that the ornament of
dark lanes in the area is formed by more distant dust concentrations.
These dark features are observable in the Rift up to Galactic latitude
of $\sim 5\degr$.  At higher latitudes distant stars of the Milky Way
probably are attenuated by the 200--250 pc clouds only.

There is no direct way to determine how deep the system of the
\object{Aquila Rift} clouds is.  As follows from the apparent
obscuration of the Milky Way and from the low-velocity CO
distribution, the \object{Aquila Rift} extends by about $20\degr$
both in Galactic longitude and latitude.  At the 225 pc distance
this corresponds to 82 pc.  This value may characterize the cloud
system depth if it is more or less spherical.  This diameter of
the cloud system fits perfectly the equation between the
logarithms of the CO line widths and the logarithms of the cloud
radii derived by Dame et al.  (\cite{dame85}, \cite{dame86}).
Consequently, if the front edge of the cloud system is at 225 pc
and the far edge at 310 pc, the center distance is at about 270
pc.  At this distance from the Sun the maximal distance of the
clouds from the Galactic plane is 70 pc ($\sim\,$$15\degr$).

At about $15\degr$ longitude the northern part of the Milky Way
(or the Central Bulge) reappears again, however it is mottled by
numerous dark lanes which are very similar to the lanes in Serpens
Cauda.  Probably these dust concentrations belong to the same
cloud system but they are nearer to the center or the far edge (up
to 310 pc from the Sun).

On the other hand, the Ophiuchus-Scorpio complex of dust and
molecular clouds between longitudes $350-0\degr$ extends from
$-7\degr$ to $+22\degr$ in latitude.  At a distance of the front
edge of the clouds of 150 pc (Strai\v zys \cite{strai84}, after
transforming to the \object{Hyades} distance modulus of 3.3 mag)
the diameter of the clouds is $\sim\,$80 pc, and the center of the
cloud system is at a distance of 190 pc.  At this heliocentric
distance the height above the Galactic plane of $22\degr$
corresponds to $\sim\,$80 pc.  Thus, both the Serpens clouds and
the Ophiuchus-Scorpio clouds reach about the same distance above
the plane.  This is an argument for a common origin of both the
\object{Aquila Rift} and the Ophiuchus-Scorpio clouds.  A
connection of both systems of molecular clouds has been suspected
by Lebrun \& Huang (\cite{lebrun}) and Dame \& Thaddeus
(\cite{dame85}) on the ground of the surface distribution and
radial velocities of CO radio emission. They have also related
this cloud system with the clouds in Lupus, on the other side of
the Galactic center.  Dame et al.  (\cite{dame87}) note that all
these local molecular clouds can be readily distinguished from the
more distant ones by their low velocity (less than 20 km/s) and a
wide extent to the northern latitudes.

According to Dame et al.  (\cite{dame87}) the mass of the
\object{Aquila Rift} clouds, estimated from the intensities of CO
lines, is about $1.5\times10^5 M_{\odot}$ at a distance of 200 pc.
However, the center of the cloud system is probably at about 270
pc, i.e., 1.35 times farther.  The new distance leads to the
increase of the surface density of gas by a factor of 1.8.  Thus
the total mass of the \object{Aquila Rift} clouds may be as large
as $2.7\times10^5 M_{\odot}$ which is  still lower than the virial
mass of the clouds calculated by Dame \& Thaddeus (\cite{dame85})
from the radial velocity dispersion (or CO line widths) and
transformed to the new distance and hence larger cloud radius.

The above conclusions about the distribution of the molecular and
dust clouds receive support from the existing Milky Way surveys in
the far infrared wavelengths.  A glance at the 240, 100 and 50
$\mu$m dust emission maps obtained by IRAS and COBE/DIRBE
(http://skyview.gsfc.nasa.gov; Schlegel et al. \cite{schlegel})
shows that the investigated area up to $b = +5\degr$ emits a
strong and uniform radiation with some intensity peaks.  The
strongest source of the far IR radiation is the complex
\object{W40} near the star \object{60 Ser}, containing a molecular
cloud, a H\ts II region and a number of point IR sources.  This
complex is considered to be unrelated with the \object{Aquila
Rift}, being at a distance of 400--700 pc (Smith et al.
\cite{smith}; Vallee \& MacLeod \cite{vallee}; Shuping et al.
\cite{shuping}). Its distance is very uncertain, being determined
mostly by the kinematical method, combining radial velocities from
molecular radio lines and the rotation curve of the Galaxy.
Reyle \& Robin (\cite{reyle}) using the Point Source Catalogue of
the DENIS infrared survey have identified a young open cluster in
the complex. The extinction $A_V$ in front of the complex is
estimated by different authors to be from 9 to 17 mag.  Since the
Galactic latitude of the complex is +3.5$\degr$ only, this
extinction originates not only in the Aquila Rift clouds.

Another somewhat fainter far IR source is the core of the
\object{Serpens molecular cloud} investigated in our Paper I. At
higher Galactic latitudes the dust and CO emission drops out and
almost vanishes at $b\geq 15\degr$.  Optically, both \object{W40}
and the \object{Serpens molecular cloud} are seen projected on a
very dark foreground created by the dust lanes of the
\object{Aquila Rift}.

The main conclusions of the investigation may be summarized in the
following items:  (1) front edge of the dark clouds in the
\object{Aquila Rift} is situated at the $225\pm55$ pc distance,
(2) the cloud complex can be about 80 pc deep, (3) the maximum
extinction in $V$ in the cloud system is close to 3.0 mag, (4) the
objects at low Galactic latitudes (like the star \object{BD --01
3542} and the cloud complex \object{W40}) exhibit much larger
extinction and are situated far behind the Rift, (5) the
\object{Aquila Rift} clouds are similar to the Ophiuchus-Scorpio
clouds:  both complexes reach more or less the same maximum height
above the Galactic plane.

\begin{acknowledgements}
      We are grateful to A. G. Davis Philip and the anonymous
      referee for important corrections.

\end{acknowledgements}

\end{document}